\documentstyle[aps,prl,multicol]{revtex}
\input {epsf.sty}
\begin{document}
\draft
\title{Superconducting Fluctuation Effects on the
	Spin-Lattice Relaxation Rate in YBa$_2$Cu$_3$O$_{6.95}$}
\author{V. F. Mitrovi{\'c}, H. N. Bachman, W. P. Halperin, M. Eschrig, J. A.  Sauls}
\address{Department of Physics and Astronomy, and
   Science and Technology Center for Superconductivity,\\
   Northwestern University, Evanston, Illinois 60208}
\author{A. P. Reyes, P. Kuhns, W. G. Moulton}
\address{National High Magnetic Field Laboratory
   Tallahassee, Florida 32310}
\date{Received 30 July 1998}
\maketitle
\begin{abstract}
We report $^{63}$Cu(2) spin-lattice relaxation rate measurements of
YBa$_2$Cu$_3$O$_{6.95}$ in magnetic fields from 2.1 T to 27.3 T obtained
from $^{17}$O(2,3) nuclear magnetic resonance spin-spin relaxation.
For $T < 120$ K, the spin-lattice rate increases with increasing magnetic 
field. We identify this magnetic field dependence with the change in
the low-energy spectral weight originating from 
d-wave pairing fluctuation corrections to the density of states.
\end{abstract}
%\narrowtext
\vspace{-11pt}
\begin{multicols}{2}
Nuclear magnetic resonance has played an important
role in   elucidating the nature of high-$T_c$ 
superconductivity\cite{rigamonti98,pennington90}. In most metals
the nuclear spin-lattice relaxation rate divided
by the temperature, $1/T_1T$, is a constant
proportional to the square of the density of states at the Fermi level, $N_F$. 
In the normal state of many high-$T_c$ superconductors an increase
in $1/T_1T$ of planar Cu with decreasing temperature has been
attributed to antiferromagnetic spin fluctuations\cite{millis90}.
At lower temperatures, in the superconducting state, the rate of planar 
Cu decreases strongly with decreasing temperature as the gap in
the quasiparticle spectrum develops.
The crossover from normal to superconducting behavior occurs around 
100 K, substantially above the transition temperature of optimally doped YBCO. 
We have investigated this cross-over experimentally and theoretically. 
We show that the cross-over can be understood quantitatively in terms of pairing fluctuation 
corrections to the spin lattice relaxation rate in optimally doped YBCO.

Because of their large anisotropy and small coherence lengths, the onset 
of superconductivity in high-$T_c$ materials is 
preceded by the effects of strong superconducting fluctuations
on the normal-state properties, including 
the specific heat\cite{annett90}, diamagnetism\cite{lee89},
nuclear spin-lattice relaxation rate\cite{kuboki89,randeria94,carretta96},
and Pauli susceptibility\cite{randeria94,bachman98b}. Here we report on the 
field dependence of  $1/T_1T$ of planar copper, $^{63}$Cu(2),  
in optimally doped YBCO. We find that below 120 K the relaxation rate 
increases with increasing field with a typical field scale of 10 T. 
We quantitatively account for this behavior in terms of 
pairing fluctuations with d-wave symmetry\cite{Eschrig98}.

Our aligned powder sample of 30\% - 40\% $^{17}$O-enriched 
YBa$_2$Cu$_3$O$_{6.95}$ has been investigated 
previously\cite{bachman98b,bachman98a,reyes97,takigawa89,hammel89}.
Our measurements cover
the temperature range 70 K to 160 K over a  wide range of magnetic fields, from 2.1 T to
27.3 T. The crystal $\hat c$-axis was aligned with the direction of the
applied magnetic field, the $z$-axis.
Low-field magnetization data show a sharp 
transition at $T_c(0)=92.5\,\mbox{K}$.
In order to study planar copper nuclear spin-lattice
relaxation we take advantage of its 
direct effect on the $^{17}$O(2,3) NMR spin-spin relaxation which we
can accurately measure using a Hahn echo sequence: 
$\pi /2$-$\tau $-$\pi $-acquire. Our typical $\pi/2$ pulse 
lengths were 1.5 $\mu $s, except at 2.1 T where pulse lengths 
were 2.5 $\mu $s, giving us a bandwidth $ > 100$ kHz.  
After the $\pi/2$ pulse the precessing nuclear spins 
dephase because of variations in
the $z$-component of the magnetic field in the sample.
The dephasing from static processes is 
recovered after the $\pi $ pulse, leaving the echo 
intensity to be determined predominantly by copper spin-lattice 
relaxation, as has been recently demonstrated\cite{recchia96}.
The $^{17}$O(2,3) \mbox{(1/2 $\leftrightarrow$ -1/2)} resonance has a 
low frequency tail owing to oxygen deficiency in a small portion of the sample\cite{reyes97}.
Its effect on our measurements can be eliminated by performing a 
non-linear least squares fit in the frequency domain for each echo, a method similar to that of
Keren {\it et al.}\cite{keren}. The success of this procedure was
checked by relaxation measurements performed on the first low frequency
satellite of the O(2,3)\cite{bachman98a}.
The oxygen resonance is much narrower than that of copper  (by a 
factor of 6 at 8.4 T) and thus $^{17}$O NMR
is more favorable for our experiments. This is 
particularly true for the high field experiments,
$H_o > 15$ T, performed in a Bitter magnet at the National High Magnetic 
Field Laboratory in Tallahassee, Florida.  The measurements for
$H_o \le 14.8 $ T were obtained with superconducting magnets.
The reliability of this technique for measuring $T_1$ of 
$^{63}$Cu(2) was tested by comparison with direct measurements of $T_1$
performed on the same sample.

We extract $T_1$ of $^{63}$Cu(2) from $^{17}$O(2,3) spin-spin 
relaxation data following the proposal of Walstedt and Cheong\cite{walstedt95} 
that the dominant mechanism for spin-echo decay of $^{17}$O is the
copper spin-lattice coupling. The $z$-component fluctuating fields from copper nuclear 
spin flips are transferred to the oxygen nuclei by Cu-O nuclear dipolar
interactions. To account for this process
Recchia {\it et al.}\cite{recchia96} derived an expression for 
the $^{17}$O spin echo height, $M(\tau )$, as a 
function of pulse spacing $\tau $,
\begin{eqnarray}
\label{eq1}
& \displaystyle
M=M_0\exp \Bigg\{ -^{17}\! \! \gamma^2 k^2 \sum_{i=1}^\nu \left[ 
\frac{^{63,65}\gamma \hbar }{r_i^3} (1-3\cos^2 \theta_i ) \right]^2 
\times & \nonumber \\
& \displaystyle
\frac{I(I+1)}{3}
\left(T_1^{(i)}\right)^2 \Big[ 2\tau /T_1^{(i)}+
& \nonumber \\
& \displaystyle
4e^{-\tau/T_1^{(i)}} - e^{-2\tau/T_1^{(i)}}-3 \Big]
- 2\tau/T_{2R} \Bigg\}.
& \! \! \! \! \! \! \! \!
\end{eqnarray}

\begin{figure}[h]
%%%%%%%%%%%%%%%%%%%   F I G U R E   %%%%%%%%%%%%%%%%%%%%
\centerline{\epsfxsize1.00\hsize\epsffile{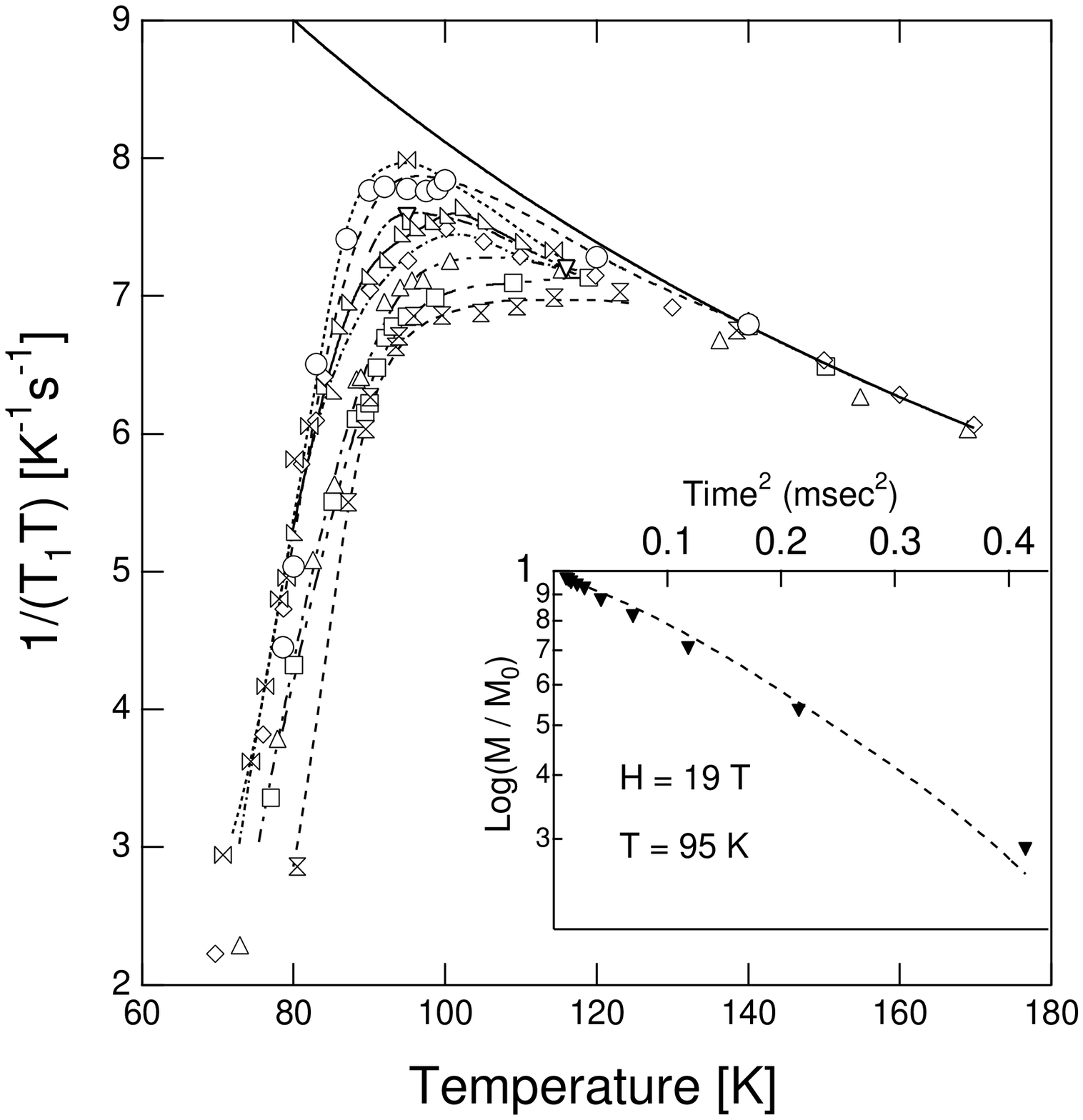}}
%%%%%%%%%%%%%%%%%%%%%%%%%%%%%%%%%%%%%%%%%%%%%%%%%%%%%%%%
\begin{minipage}{0.95\hsize}
\caption[]{\label{Fig1}\small
Spin-lattice relaxation rate of $^{63}$Cu(2) in YBCO as a function of temperature for the following
fields: 27.3 T ($\bowtie$), 22.8 T ($\bigcirc$), 19 T ($\bigtriangledown$), 
14.8 T 
($\mid \! \! \! \backslash \! \! \!_{_{\mbox{\_}}}$), 
8.4 T ($\diamond$), 5.9 T ($\triangle$), 3.2 T ($\Box$), 2.1 T 
($^{\bigtriangledown }_{\triangle }$).
Dashed lines are guides to the eye. The solid line is a fit to
$(T_1T)_{n}^{-1} \propto T_x / (T + T_x)$, $T_x=103$ K\cite{millis90}.
Inset: Spin-spin relaxation of $^{17}$O NMR at 19 T, $T=95$ K and a fit
to Eq. (\ref{eq1}).
}
\end{minipage}
\end{figure}

We performed a nonlinear least squares fit of the data to \mbox{Eq. 
(\ref{eq1})} in the range \mbox{50
$\mu $s $< \tau <$ 350 $\mu $s}, with $T_1$ of $^{63}$Cu(2) as 
a fitting parameter.  The sum was performed over all Cu neighbors in a 
radius of 12 \AA; $r_i$ is the
Cu-O distance; $\theta_i$ is the angle between the applied field 
and the Cu-O axis; $T_1^{(i)}$ is 
$T_1$ of the $i^{\mbox{th}}$ copper nucleus; $I=3/2$
is the copper nuclear spin;
$k$ is an enhancement factor due to the Cu-O 
indirect coupling which we determine to be
1.57; and $T_{2R}$ is the Redfield contribution to the rate.  An example of the 
fit is presented in the inset to \mbox{Fig. \ref{Fig1}}, at 19 T and 95 K,
and is compared with the measured relaxation profile.  
The fit to Eq. (\ref{eq1}) is 
sufficiently accurate that 
we can rely on its systematic behavior.  We have also compared our 
$1/T_1T$ data with direct measurements of $^{63}$Cu(2) spin-lattice 
relaxation rates taken from earlier 
work\cite{carretta96,hammel89,song91}  for several magnetic fields, as 
shown in \mbox{Fig. \ref{Fig3}}. 
The measurement at 7.4 T was performed on our sample\cite{hammel89}.

Our results for $1/T_1T$ are presented in \mbox{Fig. \ref{Fig1}}.  
Above  120 K there is no discernible field 
dependence within experimental accuracy of $\pm   2 \%$.
However, near the  peak in $1/T_1T$ we find that the rate increases 
with increasing magnetic field. At 95 K the rates at 
2.1 T and 27.3 T differ by 17\%. The peak in \mbox{$1/T_1T$ versus $T$} 
shifts toward lower temperature as the field increases and the rate drops sharply
in the superconducting state, consistent with reduction of $T_c$ by the
field\cite{bachman98b}.
We show below that pairing fluctuations are in quantitative agreement with
this behavior, and that a purely magnetic mechanism with a spin pseudogap
is difficult to reconcile with the field scale.

In underdoped materials the temperature dependence of the Knight shift,
$K(T)$, and the peak in
$1/T_1T$ has been associated with  the opening of a spin pseudogap
\cite{berthier96} in the spin excitation spectrum
below a temperature $T^\ast> 100\,\mbox{K}$. The temperature scale
$T^\ast$ was suggested to be
a rough measure of the pseudogap, with a magnetic field scale 
of $H^\ast=k_B T^\ast/\mu_B$, $\ge 140$ T. This exceeds by far the field
scale of $\sim 10$ T that we observe in $1/T_1T$ in our optimally doped sample. 
The large field scale, $H^\ast\gg 10$ T, for a spin pseudogap is
consistent with recent neutron scattering measurements  
that show that the resonance peak of optimally doped YBCO remains almost
unaffected in a field of 11.5 T\cite{bourges97}.

\noindent  

\begin{figure}
\begin{minipage}{0.95\hsize}
%%%%%%%%%%%%%%%%%%%  D I A G R A M S   %%%%%%%%%%%%%%%%%%%%
\centerline{\epsfxsize1.00\hsize\epsffile{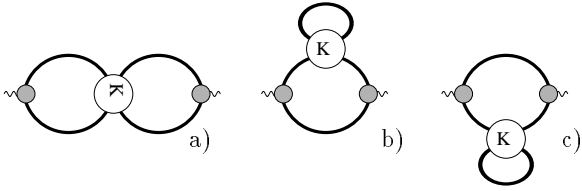}}
%%%%%%%%%%%%%%%%%%%%%%%%%%%%%%%%%%%%%%%%%%%%%%%%%%%%%%%%%%%
\begin{minipage}{0.95\hsize}
\caption[]{\label{diagrams}
Pairing fluctuation corrections, to leading order in
$T_c/E_F$, for the nuclear spin-lattice relaxation rate.
a) is the Maki-Thompson process, b) and c) are the density of states
corrections to the rate. $K$ is the impurity-renormalized
pair fluctuation propagator.}
\end{minipage}
\end{minipage}
\end{figure}

\begin{figure}[h]
%%%%%%%%%%%%%%%%%%%   F I G U R E   %%%%%%%%%%%%%%%%%%%%
\centerline{\epsfxsize1.00\hsize\epsffile{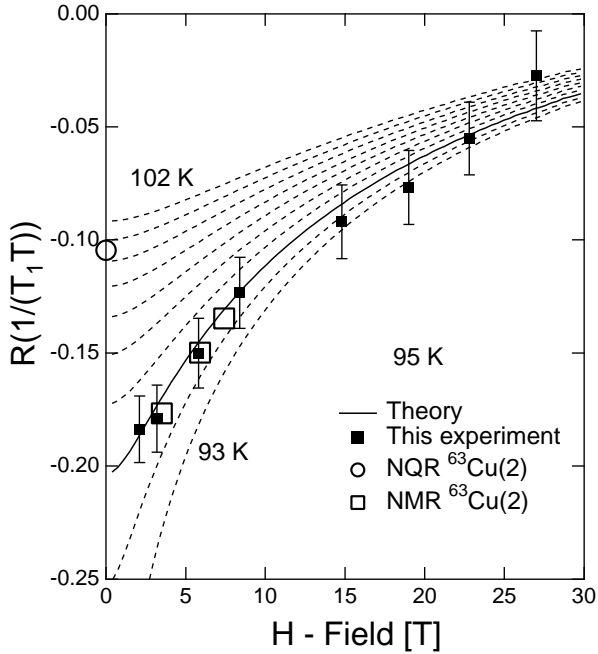}}
%%%%%%%%%%%%%%%%%%%%%%%%%%%%%%%%%%%%%%%%%%%%%%%%%%%%%%%%
\begin{minipage}{0.95\hsize}
\caption[]{\label{Fig3}\small 
Fluctuation contribution $R(1/(T_1T))$ = $\delta (T_1T)^{-1}$ / 
$(T_1T)^{-1}_{n}$ of $^{63}$Cu(2)
spin-lattice relaxation rate as a function of magnetic field at 95 K. 
The dashed curves are d-wave calculations for temperatures ranging from 93 
to 102 K in increments of 1 K. The solid curve is calculated for 95 K. The open 
circle is $R(1/(T_1T))_{NQR}$ at 95 K\cite{song91}. The open squares are from 
direct measurements of the $^{63}$Cu(2) $T_1$ at 3.5 T by Y.-Q. 
Song\cite{song91}, 5.9 T by Carretta {\it et al.}\cite{carretta96}, 
and 7.4 T by Hammel {\it et al.}\cite{hammel89}.}
\end{minipage}
\end{figure}

In high-$T_c$ materials  superconducting fluctuations are expected to have
a significant effect on $1/T_1$ near $T_c$. 
Diamagnetic fluctuations do not play a role in our measurements of $T_1$ 
since they alter the magnetic field mainly along the axis parallel to the
applied field; only transverse fields contribute to relaxation of
the $\it z$-component of the nuclear spin. The pairing fluctuation contributions to
the rate result from fluctuation corrections to the 
density of states (DOS) and from the Maki-Thompson (MT) 
corrections
to the local dynamical susceptibility. The corresponding Feynman
diagrams for these corrections are 
 shown in Fig. \ref{diagrams}.
The propagators and vertices are defined below and in Ref.\cite{Eschrig98}.
A systematic analysis in $T_c/E_F\ll 1$, where $E_F$ is 
the  Fermi energy,
shows that all other pairing fluctuation contributions are negligible
for local quantities like $1/T_1$\cite{Eschrig98}.
The pairing fluctuation correction is sensitive to the symmetry
of the order parameter fluctuations because of the difference in sign of the MT
(positive) and DOS (negative) corrections, and because of the sensitivity
of the non-s-wave pairing fluctuations to disorder.
In the case of s-wave pairing
fluctuations the dominant contributions to the rate come from 
the positive MT processes\cite{kuboki89}, which are insensitive
to non-magnetic disorder. A magnetic field suppresses the MT and
DOS contributions, and leads to a suppression of the rate for
s-wave. In the case of d-wave pairing the field dependence of $1/T_1$
is reversed compared to that for s-wave pairing. Scattering
by non-magnetic disorder 
\noindent 
leads to strong suppression of the MT
corrections for d-wave fluctuations. The DOS corrections survive
non-magnetic scattering, but are suppressed by a magnetic field leading
to an increase in $1/T_1$ with increasing field,
even for modest levels of disorder. As we show below, our results
provide a consistent and quantitative account of the field dependence
of the nuclear spin-lattice relaxation rate above $T_c$.

In order to make a quantitative comparison between the leading order pairing
fluctuation corrections and the 
experimental field dependence of the rate
we isolate the 
fluctuation corrections to the experimental rate by
writing $(T_1T)^{-1}_{tot}= (T_1T)^{-1}_{n} + \delta (T_1T)^{-1}$,
where the normal-state rate is fit to the AFM Fermi-liquid
model\cite{millis90}, $(T_1T)^{-1}_{n} \propto T_x / (T + T_x)$.
We obtain $T_x = 103$ K from a fit to high temperature data at 8.4 T.
The fluctuation contributions are indicated by
$\delta (T_1T)^{-1}$. These values,  normalized by
$(T_1T)^{-1}_{n}$, are plotted in \mbox{Fig. \ref{Fig3}}
 as a function of magnetic field at 95 K along with our theoretical calculations 
of the pairing fluctuation corrections.

The calculations of $1/T_1$ assume a quasi-2D
cylindrical Fermi surface, with an isotropic in-plane
Fermi velocity $\vec v_f$.
We expect the pairing fluctuations to be predominantly two-dimensional
in a magnetic field because of Landau-level quantization. A cross-over
to 3D fluctuations is possible below $\sim 2$ T near $T_c$.
A summary of the calculation is provided here;
more details can be found in Ref. \cite{Eschrig98}.
The pairing interaction is $V(\vec p,\vec p\,')
=\eta(\vec p)\cdot g\cdot\eta(\vec p\,')$, where
$\eta(\vec p)$ is the normalized pairing amplitude; for s-wave
pairing $\eta(\vec p)=1$ while for d-wave pairing
$\eta(\vec p)=\sqrt{2}\cos 2\psi$, where $\psi$ is the angle 
between the crystallographic $\hat a$-axis and $\vec{p}$.

The pair fluctuation propagator is defined in terms of the sum over
ladder diagrams in the particle-particle interaction channel;
the propagator factorizes into $\eta(\vec p)L(Q)\eta(\vec p\,')$, where
$ L(Q)^{-1} = g^{-1}-T\sum_{\epsilon_n} B_2(\epsilon_n,Q)$,
$B_2(\epsilon_n,Q)= \sum_{\vec p} \eta (\vec p)
\tilde\eta(P,Q) G(P)G(Q-P)$ and $G(P)$ is the quasiparticle
Green's function. We use a short-hand notation:
$P\equiv(\epsilon_n,\vec p)$,
$P'\equiv(\epsilon_{n'},\vec p\,')$ for fermion quasiparticles,
and $Q \equiv(\omega_l,\vec q)$ for bosonic Matsubara energy
and pair momentum of the fluctuation modes; the pair momentum,
$\vec{q}$, is quantized because of orbital quantization in a magnetic field.
We include disorder via the standard averaging procedure for dilute
impurity concentrations\cite{Abgo}. Impurity scattering introduces an
elastic scattering time in the quasiparticle Green's function,
$G(P)=(i\epsilon_n-\Sigma(\epsilon_n)-\xi(\vec p))^{-1}$,
where $\xi(\vec p)=\epsilon(\vec p)-\mu$ is the quasiparticle excitation 
energy, $\Sigma(\epsilon_n )=-(i/2\tau +i/2\tau_\phi)\mbox{sign}(\epsilon_n)$
is the self energy, and $\tau$ is the elastic scattering lifetime.
We include inelastic scattering through the lifetime $\tau_\phi$.
Impurity scattering modifies the fluctuation propagator
directly through a vertex correction in the particle-particle channel,
$\tilde\eta(P,Q)
=\eta(\vec p)+\sum_{\vec p}\eta(\vec p)G(P)G(Q-P)C(\epsilon_n,Q)$,
where $C(\epsilon_n,Q)^{-1}=\tilde\alpha^{-1}-\sum_{\vec p}G(P) G(Q-P)$
is an impurity Cooperon-like propagator and
$\tilde\alpha=1/2\pi\tau N_F$ is the impurity scattering vertex.
The full impurity-renormalized pair propagator, $K(P,P',Q)$,
which enters the dynamical susceptibility diagrams shown in
\mbox{Fig. \ref{diagrams}}, is given
by $\tilde\eta(P,Q)L(Q)\tilde\eta(P',Q)$.
The leading order fluctuation correction to $1/T_1$
then follows from the Feynman rules for evaluating the diagrams\cite{Abgo}
and is given by,
\begin{eqnarray}
\label{MAKI}
\delta \chi_{M}(\omega_m)&=&
-2| \vec{A\, } |^2 \sum_{n,Q} B_1(\epsilon_n,Q)
B_1(\epsilon_n-\omega_m,Q) L(Q),
\nonumber
\\
\label{DOS}
\delta \chi_{D}(\omega_m)&=& 4| \vec{A\,}|^2
\sum_{n,Q} G_1(\epsilon_n-\omega_m) 
\frac{\delta B_2(\epsilon_n,Q)}{\delta \Sigma(\epsilon_n)}L(Q),
\\
\label{rate}
\delta (T_1T)^{-1}&=&
 \lim_{\omega \to 0} 2\,\mbox{Im}
\frac{\delta \chi_{M}(\omega )+\delta \chi_{D}(\omega )}{\omega}\,,
\end{eqnarray}
with $B_1(\epsilon_n,Q)=\sum_{\vec p} \tilde \eta(P,Q) G(P) G(Q-P)$,
$G_1(\epsilon_n)=\sum_{\vec p}G(P)$, and $|\vec{A\,}|^2$ are 
momentum-averaged hyperfine form factors\cite{Eschrig98}.
We analytically continue Eqs. (\ref{DOS}) to real energies
using Eliashberg's technique\cite{eliash} to obtain
$\delta \chi_{M}(\omega )$ and $\delta \chi_{D}(\omega)$.
The zero frequency limit in Eq. (\ref{rate}) is performed 
analytically and the resulting equations are evaluated numerically.
The sum over $Q$ includes a summation over all Landau levels and
over all dynamical fluctuation modes,
in order to extend the theory beyond the region of validity for
static, long-wavelength fluctuations, and beyond the 
lowest-Landau-level approach.

The experimental zero-field transition temperature of 92.5 K 
determines the temperature scale for the theoretical calculations.  
The mean-field transition temperature, $T_c$(8.4 T) = 80.9 $\pm.3$ K, 
which is determined by the divergence of the pair fluctuations, 
is obtained from our fit to spin susceptibility\cite{bachman98b}.
We assumed $\hbar/2\pi\tau_{\phi}=0.02k_BT_c$ and
$\hbar/2\pi\tau=0.2k_BT_c$, and
there is one fitting parameter for the overall scale of the 
fluctuation contributions to $1/T_1$.
Our theoretical calculation for the field dependence of the 
fluctuation correction  is shown in
Fig. \ref{Fig3} for d-wave pairing. The rate increases because of the
suppression of the (negative) DOS contribution to the rate
by the magnetic field. The results agree quantitatively with the 
experimental data at $T=95\,\mbox{K}$ and provide strong
evidence for d-wave pairing fluctuations. For s-wave pairing 
the calculated rate (not shown) {\it decreases} with increasing magnetic 
field because of the suppression of the (positive) MT term. 
\noindent 

Carretta {\it et al.}\cite{carretta96} reported experimental 
evidence for a positive contribution to the rate that was attributed to
the MT process.
These authors compared  nuclear quadrupolar resonance (NQR)
relaxation measurements and NMR relaxation at 5.9 T and found an
NQR rate that is higher than the NMR rate in a range 
of $\sim 10$
K above $T_c$, a result which is similar to our NQR measurement
shown in \mbox{Fig. \ref{Fig3}}. They interpret the decrease
from the higher NQR to the lower NMR rate at 5.9 T in terms of
s-wave pairing fluctuations, which implies a dominant MT term.
However, our data in \mbox{Fig. \ref{Fig3}} shows that there is
no significant MT contribution to the 
NMR rate at fields above 2.1 T.
Our analysis of the field dependence of the data is in excellent
agreement with the 
  theory of d-wave pairing fluctuations, and 
disagrees with the theory based on s-wave fluctuations.
Possible explanations for the apparent discrepancy  between the NQR
rate and the low-field NMR rate include 
an admixture of s-wave and d-wave fluctuations
induced by orthorhombic anisotropy\cite{Eschrig98}, and 
the 2D to 3D crossover regime at low fields.

In summary, we have determined the $^{63}$Cu(2) spin-lattice relaxation rate as a function 
of magnetic field from 2.1 T to 27.3 T. We found that $1/T_1T$
increases with increasing field in the temperature range $T < 120$ K,
which we can account for quantitatively with the theory of
d-wave pairing fluctuations in 2D. Our results are consistent with
d-wave pairing in YBCO, and inconsistent with dominant 
s-wave pairing. We found that the characteristic field
scale for the suppression of the fluctuation corrections, $\delta(T_1T)^{-1}$, 
is $\sim 10$ T,
which is  an order of magnitude smaller than the expected field scale for
a purely magnetic scenario for the pseudogap.

We gratefully acknowledge useful discussions with Y. -Q. Song, D. Rainer, 
D. Morr, M. Randeria.  We are particularly thankful to C. Hammel
for providing the sample. This work is
supported by the  National Science Foundation (DMR 91-20000) through the Science and 
Technology Center for Superconductivity. The work at the National High
Magnetic Field Laboratory was supported by the National Science Foundation 
under Cooperative Agreement No. DMR95-27035 and the State of Florida.
ME also acknowledges support from the Deutsche Forschungsgemeinschaft.
\vspace{-0.5cm}
\bibliographystyle{unsrt}

\vspace{0.5cm}
\end{multicols}
\end{document}